# Driving LED in a Nanosecond Regime by a Fast Operational Amplifier

**Joachim Rose, Stella Bradbury, Isabel Bond, Paul Ogden, Andrew Price, Richard Oliver and Yerbol Khassen**

**Abstract**

It is widely believed that the generation of high speed optical signals is not the job for an LED. However this work is done to show that there are techniques which can be used to produce nanosecond square pulses from a diode. Rise and fall times of a typical 10ns long signal were 1-2 ns and the intensity of the emission could be controlled by the supply voltage. The wavelength of the radiation was 472 nm, which is blue in colour, but any longer or even shorter wavelengths can similarly be used. The consistency of the experiment and its theoretical model was analysed by computer simulations using OrCAD and PSPICE.

## Introduction

An LED consists of a semiconductor pn junction where electrons recombine with holes when forward biased and power dissipated in the form of light. The frequency of light depends on the band gap of the semiconductor element, which is usually based on InGaN (indium gallium nitride). The blue LED was chosen to be used due to the design of a particular application although longer wavelengths are easier to handle because of the smaller band gap. Current flows easily from the p-side (anode) to the n-side (cathode) but not in the opposite direction. The reason is that the electrons are in energetically higher state of the conduction band so they try to obtain the ground state by reaching the valence band. The blue LEDs are less efficient than the red or green LEDs and it causes the leakage in the form of thermal energy which may result in hot temperatures across the junction of about 150°C. Figure 1 illustrates the processes in the diode with potentials applied.

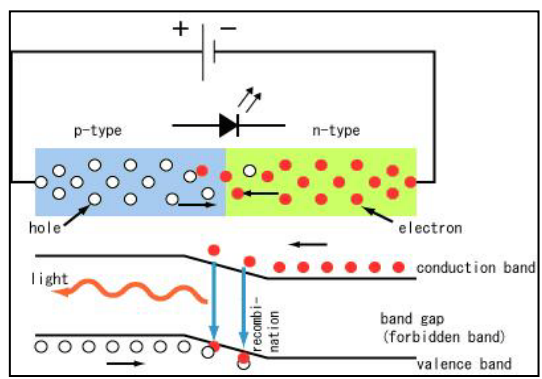

Figure 1    Schematic diagram of a circuit with a diode where electrons combine with holes and produce output emission. Below is the process of recombination in the junction; the larger the band gap the shorter the wavelength of the photon emitted [5].

When forward biased, the ideal diode is simply a short circuit and when reverse biased, an open circuit. According to Veledar (2007), the equivalent circuits of a diode further classified into DC (up to ~10 kHz) and AC models (>10 kHz), because higher frequencies result in new parameters such as parasitic capacitance and resistance. The model of the diode for this experiment is called *large signal* and it has a schematic form drawn in Figure 2, but there are also models called *small bias* and *large bias*. Massobrio and Antognetti (1993) explained that the aspect of charge storage must be considered in large signal model, because 'if there was no charge storage, the devices would be infinitely fast as there would be no charge inertia and the currents could be changed in zero time'. The *diode capacitance* $C_D$ is the sum of two capacitances: the *junction capacitance* $C_j$ and the *diffusion capacitance* $C_d$ given by

$$C_D = \frac{dQ_D}{dV_D} = \begin{cases} \tau_D \frac{dI_D}{dV_D} + C_j(0)\left(1 - \frac{V_D}{\varphi_0}\right)^{-m} & \text{for} \quad V_D < FC \times \varphi_0 \\ \tau_D \frac{dI_D}{dV_D} + \frac{C_j(0)}{(1-FC)^{1+m}}\left(1 - FC(1+m) + \frac{mV_D}{\varphi_0}\right) & \text{for} \quad V_D \geq FC \times \varphi_0 \end{cases}$$

(1)

with m being equal to 0.33 for a linearly graded junction or 0.5 for an abrupt junction and FC being a factor between 0 and 1. Detailed meanings of the terms can be found in Appendix. The junction capacitance is due to the dopant concentrations in the depletion region which is dominant term for reverse and small forward biases. For moderate forward and beyond, the diffusion capacitance dominates due to the minority-carrier charges injected into the neutral regions and effects of $C_j$ are therefore neglected. It is assumed that the temperature does not contribute to the operation of the real diode, because the pulse periods (~1 ms) were used are about 100,000 times larger that the pulse width and the diode is off almost all the times, i.e. the duty cycle is low. However, the temperature modifications are needed when considering frequencies below 10 kHz.

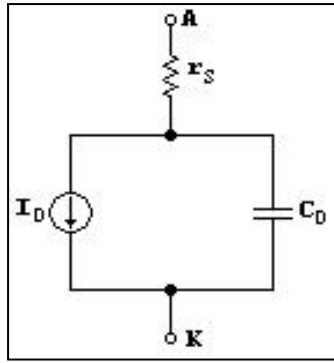

Figure 2    Large signal model for an ideal diode, where ideal diode has parasitic resistance in series and capacitance in parallel connected to anode A and cathode K.

Another main electrical characteristic of a diode is derived from the experimentally determined current-voltage relationship. Veledar (2007) assumed that 'the measured current is purely resistive' and this 'relationship is established at steady state'. The model parameters have been obtained from the curve fittings, in other words, the theoretical expressions for the LED characteristics were modified to fit the experimental data and it is simply because there is no identical LED in the world with the same parameters. Mathematically, the diode is modelled by a nonlinear current source $I_D$ according to the following equations:

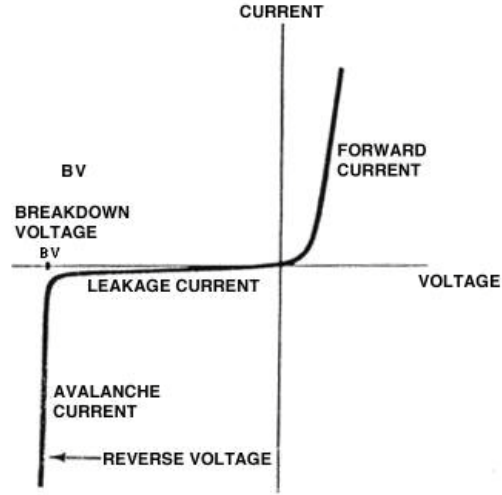

Figure 3   Theoretical model of I-V relationship of the real diode.

$$I_D = \begin{cases} I_S\left(e^{\frac{eV_D}{nkT}} - 1\right) + V_D GMIN & \text{for} \quad -5\frac{nkT}{e} \leq V_D \leq 0 \\ -I_S + V_D GMIN & \text{for} \quad -BV < V_D < -5\frac{nkT}{e} \\ -IBV & \text{for} \quad V_D = BV \\ -I_S\left(e^{\frac{e(BV+V_D)}{kT}} - 1 + \frac{eBV}{kT}\right) & \text{for} \quad V_D < BV \end{cases}$$

(2)

GMIN is the conductance (opposite to resistance) and it cannot be equal to zero. Finally, these equations have to predict the relationship between current and voltage across the diode and result in the following diagram (Figure 3).

For the purposes of the experiment the reverse voltage has not been studied because the work was only concentrated on producing flashes of light rather than achieving a complete spectrum of the LED profile system.

**Methods**

Methods included software simulations on a PC program called *PSPICE*. PSPICE is a version of SPICE program (PSPICE is an acronym for Personal Simulation Program with Integrated Circuit Emphasis). Most manufacturers supply SPICE models for their products and devices so it is easier to analyse the theoretical models in computer simulations. PSPICE algorithms for large signal model of a diode are the same for earlier version of the program called SPICE2[4]. The full code of the circuit in PSPICE is given in Appendix.

Initially the diode had to be prepared for the experiment and its parameters should be determined. By applying DC current with voltages between 0 to 3.5V through an LED and having a resistor in series generated the voltage-current relationship, where voltage drop across the LED was related to the current flow through the resistor. The parasitic resistance $r_S$ can be determined by:

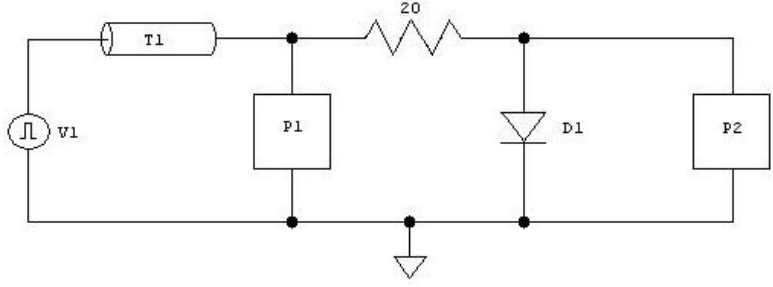

Figure 4    The diode is connected in series with a 20 ohm resistor and active probes P1 and P2 are attached as indicated in the diagram. T1 is a coax cable which transmits the pulses from the signal generator V1.

$$I\frac{dV}{dI} = r_S I + \frac{nkT}{e}$$
$$\Rightarrow y = kx + b \qquad (3)$$

which is a linear graph with a slope $r_S$. The emission coefficient is calculated from the value of the point of intersection with y axis. Saturation current $I_S$ is possible to obtain if natural logarithm of I placed against $\frac{e(V - r_S I)}{nkT}$, so that crossing point with lnI axis is $\ln I_S$.

$$I = I_S e^{\frac{V - r_S I}{nkT}} \qquad (4)$$

Furthermore, the following circuit has been analysed for capacitance parameters and the outputs of both experimental measurements and SPICE model was compared for validity (Figure 4). Square pulses were sent with a rise and fall times of 800ps and pulse width of 10ns every 1ms. Voltage amplitude varied from 1 to 3.8 volts. It was important to take into account the internal capacitance and inductance of the coax cable as well as two probes because high frequencies could significantly alter the outcome.

From the shape of the pulses seen on the oscilloscope, the junction capacitance could be found at small forward biases whereas effects of the diffusion capacitance at large biases. The value of the 20 ohm resistor was chosen to cancel the internal impedance of the coax cable which is given by

$$Z = \sqrt{\frac{L}{C}} \qquad (5)$$

and the reflection coefficient is the lowest if the load and impedance is equal ($Z = Z_I$).

$$\varepsilon = \frac{Z - Z_I}{Z + Z_I} \qquad (6)$$

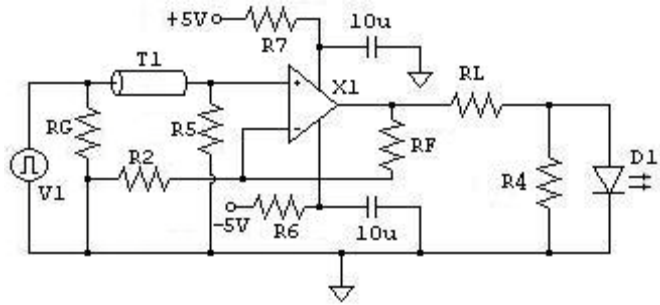

Figure 5    Fast LED driver using operational amplifier with the gain of Vout/Vin = (R1+RF)/R1. Vin has a maximum value of 1V so that the gain should be roughly more than 2.5 in order to light up the diode. The voltage sources have to be set to ±5V, according to the data sheet of the op amp. The circuit was analysed in PSPICE by the following model:

*Operational Amplifier*

Following circuit (Figure 5) was used to drive the LED using fast operational amplifier AD8009[2][3], whose spice model can be found on the manufacturer's website.

## Results

The measurements of the diode I-V relationship has been performed and it reasonably confirmed the predicted model

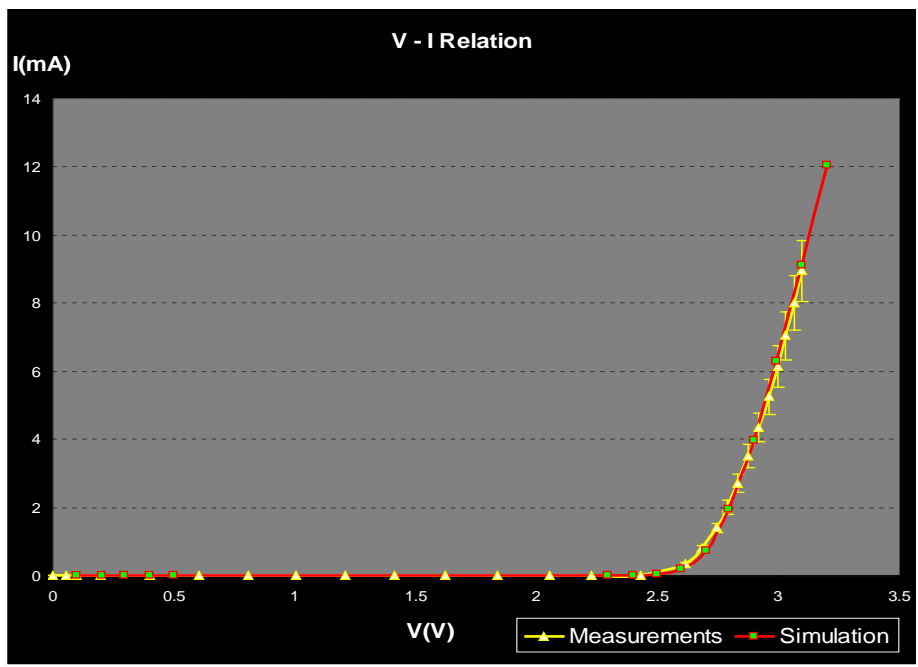

Figure 6.    Measurements and simulation results nearly coincide on V-I graph confirming the model. This was obtained by DC sweep analysis in PSPICE.

Slope $r_S$ has been calculated by plotting conduction * current versus current.

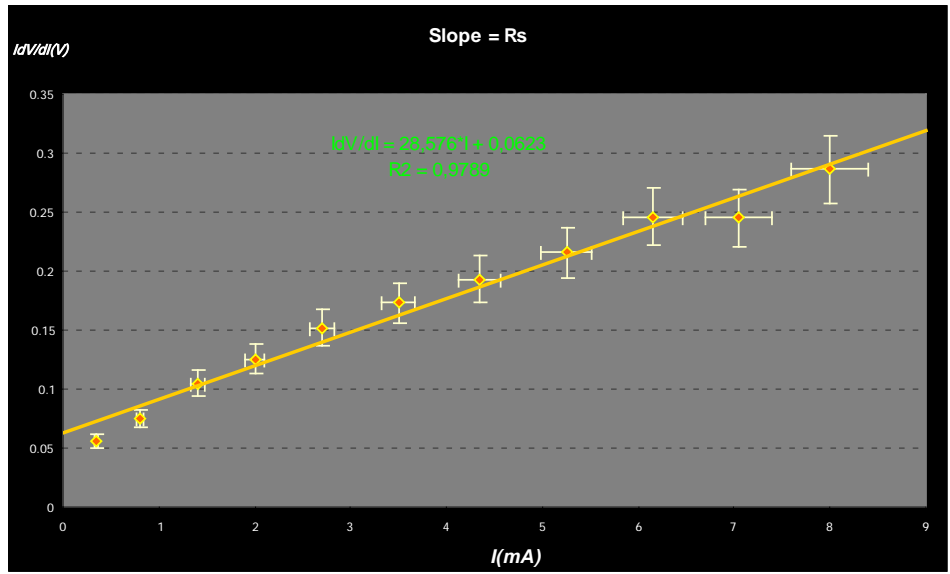

Figure 7. $r_S$ was determined as $29 \pm 4$ Ohms and the emission coefficient n was found as 2.42.

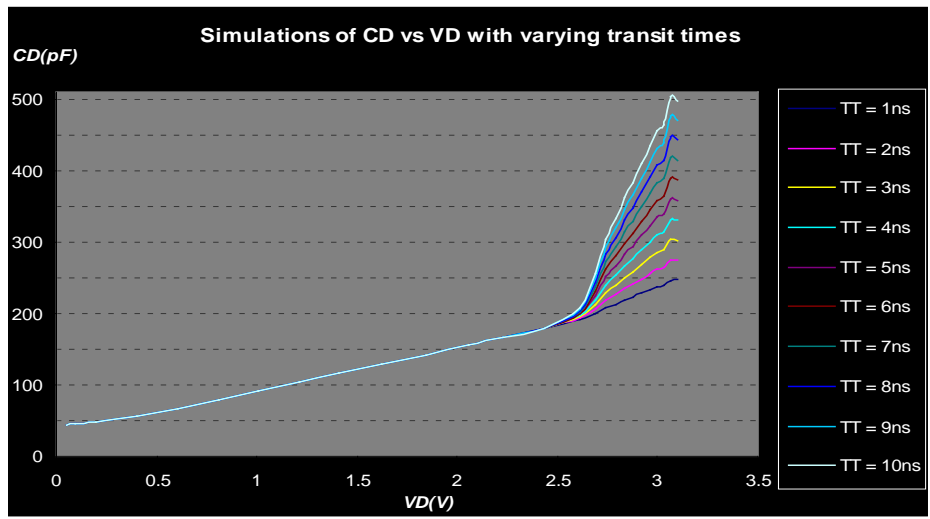

Figure 8  This plot provides evidence that the diffusion capacitance takes over at $V_D \geq FC \times \varphi_0$ and the transit time significantly affects the total capacitance. TT for the LED was found as 2.85ns.

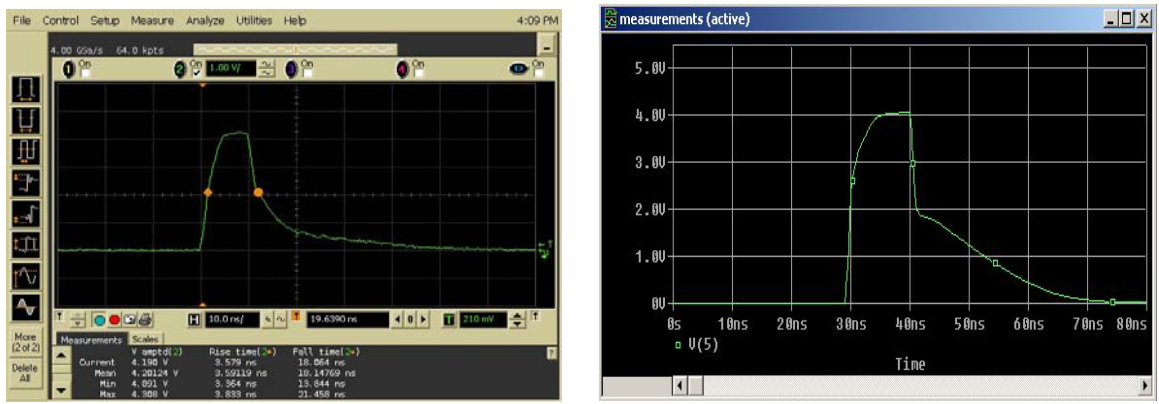

Figure 9  The above diagram is actual measurement of voltage change of the circuit illustrated in Figure 4. The orange dots indicate the rise and fall time measurements, which are given just below the graph. The below plot is the corresponding PSPICE simulation with helped to identify the parameters of the LED.

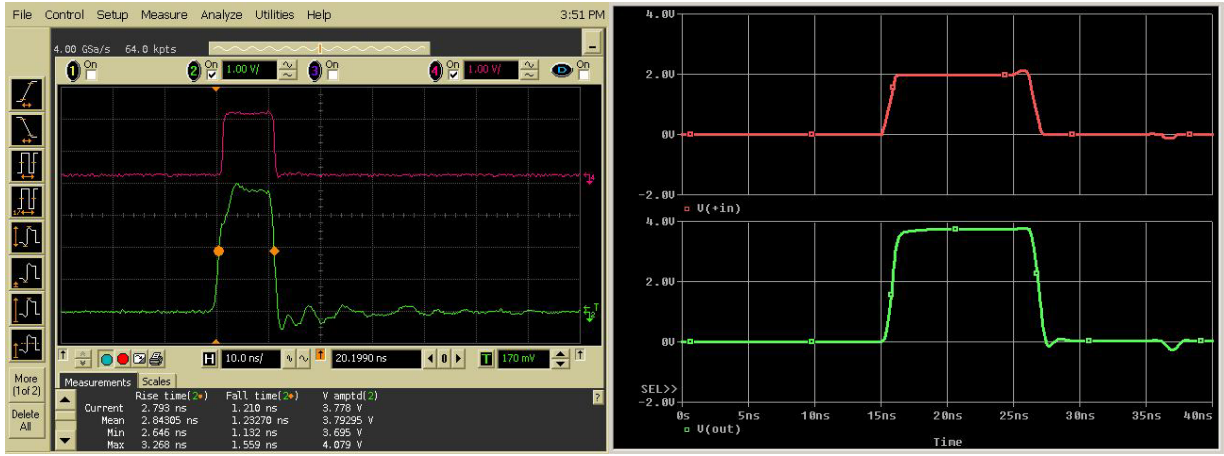

Figure 10    The red line is the 2V input and the green one is the output voltage of the LED with the gain of two. The circuit is described in Figure 5 where the feedback resistor is 200 Ohms. The rise and fall times in the simulation are 2.3ns and 1.1ns respectfully and they were verified by the measured values on the oscilloscope.

## Conclusion

This paper covered the development of electronic circuits capable of generating fast optical pulses off an LED. One big advantage if the device is extremely fast response to external signals. Rise and fall times of 1 ns have been achieved using the fast operational amplifier. Further improvements of the circuit may result in even faster scale of below than a nanosecond. One possibility is an application of three amplifiers joined together in a triangle. The lion's share of the work was concentrated on the exact modelling and measuring actual parameters of the circuit components such as the diode. Even though the LED worked as predicted, the limitation were the unknown parameters of the diode and more detailed study of its behaviour should be modelled.

The study concluded that the capacitive characteristics of the LED must be determined accurately as it significantly impacts the function of the semiconductor system and the circuit as a whole. The device is more successful to work with small voltage signals (~200mV) where either the switching speed of the amplifier increases or the gain factor rises up to 10. The last option is more preferable because the LED threshold is around 2.5 V and the necessary voltage amplification must be obtained. There are number of applications for this device, for example it is suitable for the research in neutrino detection where it provides the necessary calibration of scintillation counters. Today the neutrino experiment is the hot topic in cosmology and particle physics where the sensitivity of devices such as the one described in this paper are pushed to the limit.

## References

1. Agilent Technologies, Inc. (2003). Agilent T-13/4 (5 mm) Precision Optical Performance InGaN Blue and Green Lamps, *Data Sheet*. http://octopart.com/info/Avago/HLMP-CB15-P0000

2. Analog Devices. AD8009: 1 GHz, 5,500 V/μs Low Distortion Amplifier: *Data Sheet*. Available online at: http://www.analog.com/static/imported-files/data_sheets/AD8009.pdf

3. Analog Devices. AD8009 SPICE Macro-Model Rev B SMR/ADI, 8/97. Available online at: http://www.analog.com/Analog_Root/static/techSupport/designTools/spiceModels/license/spice_general.html?cir=ad8009.cir

4. Massobrio, G. & Antognetti, P. (1993). Semiconductor Device Modeling with SPICE (2nd ed.). New York: McGraw Hill

5. NeaEnergo. What is an LED? Available online at:
http://www.neaenergo.com/what-is-an-led

6. Veledar, O. (2007). Development of Nanosecond Range Light Sources for Calibration of Astroparticle Cherenkov Detectors. Newcastle: University of Northumbria

7. Veledar, O. et al.(2006). Simple techniques for generating nanosecond blue light pulses from light emitting diodes. Newcastle: Northumbria University

8. The SNO Collaboration (1999). The Sudbury Neutrino Observatory. arXiv:nucl-ex/9910016v2

## Appendix

General Form         D<name><(+)node><model name>[area value]

Model Form     .MODEL <model name> D[model parameters]

Arguments and Options
                 <(+)node>        the anode
                 <(-)node>        the cathode

```
DIODE MODELLING
.SUBCKT PROBE a d
Ra a b 215
La b c 22nH
Lb c d 22nH
Ca b 0 0.18pF
Cb c e 0.42pF
Rb e 0 100
Cc d f 0.6pF
Rc f 0 15
.ENDS
V1 1 0 7.6V
I1 2 0
+PULSE 152mA 0 25ns 1ns 1ns 10ns 1ms
D3 5 0 D1
.MODEL D1 D(CJO=43E-12 RS=28.58 DEV=5% N=2.42 TT=2.85E-9 EG=1.71
+FC=0.5 VJ=1 M=0.5 IS=4.26E-20)
RD 1 2 50
R1 3 5 20
X1 3 4 PROBE
X2 5 6 PROBE
R2 6 0 100k
R4 4 0 100k
* CABLE
T1 2 0 3 0 LEN=1 R={78u*SQRT(S)} L=250nH
+G={1.6p*S} C=100pF
.TRAN 10ps 100ns
```

```
          .PROBE
          .END

AD8009 SPICE model of the fast driver        Rev B SMR/ADI 8-21-97
VDC a 0 DC 0
+PULSE 0 2V 10NS 1ns 1ns 10ns 1ms
RG 0 a 50
VDC+ m 0 5V
VDC- n 0 -5V
RF c b 200
R2 0 b 200
RL c out 1
R4 out 0 100K
R5 +in 0 50
R6 f n 1
R7 e m 1
C1 e 0 10uF
C2 f 0 10uF
D3 out 0 D1
.MODEL D1 D(CJO=43E-12 RS=28.58 DEV=5% N=2.42 TT=2.85E-9 EG=1.71
FC=0.5 VJ=1 M=0.5 IS=4.26E-20)
T1 a 0 +in 0 LEN=1 R={78u*SQRT(S)} L=250nH
+G={1.6p*S} C=100pF
.SUBCKT OpAmp 1 2 99 50 28
* The source code of sub-circuit file can be found in the reference [3].
X1 +IN b e f c OpAmp
```

PSPICE Diode Model

| Symbol | .MODEL statements in PSPICE | Description | Values used | Default value in PSPICE |
|---|---|---|---|---|
| $I_S$ | IS | Saturation current | 2.46e-20 | 1e-14 |
| n | N | Emission coefficient | 2.42 | 1 |
| $\tau_D$ | TT | Transit time | 2.85e-9 | 0 |
| $C_j(0)$ | CJO | Zero-bias junction capacitance | 43e-12 | 0 |
| $\varphi_0$ | VJ | Junction potential | 1 | 0 |
| m | M | Grading coefficient | 0.5 | 0.5 |
| FC | FC | Coefficient for forward-bias depletion capacitance formula | 0.5 | 0.5 |
| $r_S$ | RS | Parasitic resistance | 28.58 | 0 |
| $dI_D$ | | Change in current across the diode | | |
| $V_D$ | | Voltage across the intrinsic diode only | | |
| $dV_D$ | | Change in voltage drop across the diode | | |
| $E_g$ | EG | Energy bandgap | 1.71 | 1.11 (eV) |
| BV | BV | Breakdown voltage | Default | Infinite |
| IBV | IBV | Reverse breakdown knee current | Default | 1e-10 |
| GMIN | | Conductance (unit is 'mho') | Default | 1e-12 |